\documentclass[12pt,a4paper]{article}
\pdfoutput=1
\usepackage{color}
\usepackage{amssymb,amsmath,bm,bbm}
\usepackage{epsf}
\usepackage{epsfig}
\usepackage{afterpage}
\usepackage{longtable}
\usepackage[dvipsnames]{xcolor}
\usepackage[linktoc=page,bookmarks=false,colorlinks=false,
linkbordercolor=RoyalBlue,citebordercolor=ForestGreen,urlbordercolor=CornflowerBlue]{hyperref}
\usepackage{latexsym,mathrsfs,dsfont}
\usepackage[normalem]{ulem} 
\usepackage[compress]{cite}
\usepackage{graphicx}
\usepackage{url}
\usepackage{paralist}
\usepackage{bbold}

\newcommand{\ord}{\mathcal{O}}

\newcommand{\IM}{{\rm Im}}
\newcommand{\RE}{{\rm Re}}

\newcommand{\gev}{\, {\rm GeV}}

\newcommand{\vcb}{|V_{cb}|}

\newcommand{\vub}{|V_{ub}|}

\newcommand{\bsi}{B_6^{(1/2)}}
\newcommand{\bei}{B_8^{(3/2)}}

\def\epe{\varepsilon'/\varepsilon}
\newcommand{\beq}{\begin{equation}}
\newcommand{\eeq}{\end{equation}}
\newcommand{\be}{\begin{equation}}
\newcommand{\ee}{\end{equation}}
\newcommand{\bi}{\begin{itemize}}
\newcommand{\ei}{\end{itemize}}
\newcommand{\ba}{\begin{array}}
\newcommand{\ea}{\end{array}}
\newcommand{\beqa}{\begin{eqnarray}}
\newcommand{\eeqa}{\end{eqnarray}}
\newcommand{\bea}{\begin{eqnarray}}
\newcommand{\eea}{\end{eqnarray}}
\newcommand{\beqn}{\begin{eqnarray}}
\newcommand{\eeqn}{\end{eqnarray}}

\definecolor{red}{cmyk}{0,1,1,0.4}

\def\kpn{K^+\rightarrow\pi^+\nu\bar\nu}
\def\klpn{K_{L}\rightarrow\pi^0\nu\bar\nu}

\usepackage{fancyhdr}
\advance \headheight by 3.0truept       

\begin{document}


\vspace{-14mm}
\begin{flushright}
        {AJB-18-9}
\end{flushright}

\vspace{8mm}

\begin{center}
{\LARGE\bf
\boldmath{$\epe$-2018: A Christmas Story}
\\[8mm]
{\large\bf Andrzej~J.~Buras \\[0.3cm]}}
{\small 
      TUM Institute for Advanced Study, Lichtenbergstr.~2a, D-85748 Garching, Germany\\
Physik Department, TU M\"unchen, James-Franck-Stra{\ss}e, D-85748 Garching, Germany\\  E-mail: aburas@ph.tum.de }
\end{center}

\vspace{4mm}

\begin{abstract}
\noindent
I was supposed to review the status of $\epe$ both at the CKM Workshop in September in Heidelberg and recently at the Discrete 2018 Conference in Vienna. Unfortunately I had 
to cancel both talks for family reasons. My main goal in these talks was
to congratulate NA48 and KTeV collaborations for the discovery of new sources 
of CP violation through their heroic efforts to measure the ratio $\epe$ in the
1980s and 1990s with final results presented roughly 16 years ago. As I will not attend any other conferences this year I will reach this goal
in this writing. In this context I will give  arguments, why I am 
convinced about the presence of new physics in $\epe$ 
 on the basis of my work with Jean-Marc G{\'e}rard within the context of the Dual QCD (DQCD) approach and why  RBC-UKQCD lattice QCD collaboration and in particular  Chiral Perturbation Theory  practitioners are still unable to reach this conclusion. I will demonstrate that even in the presence of pion loops, as large as 
advocated recently by Gisbert and Pich, the value of $\epe$ is significantly below the data, when 
the main non-factorizable QCD dynamics at long distance scales, represented 
in DQCD by {\em the meson evolution}, is taken into account.
As appropriate for a Christmas story, I will prophesy
 the final value of $\epe$ within the SM, which should include in addition to the correct matching between long and short distance contributions, isospin breaking effects, NNLO QCD corrections to both QCD penguin and electroweak penguin 
contributions and final state interactions. Such final SM result will probably 
 be known from lattice QCD only in the middle of the 2020s, but already in 2019 we should be able to see some signs of NP in the next result on $\epe$ from RBC-UKQCD. In this presentation I try to avoid, as much as possible, the overlap 
with my recent review of Dual QCD in \cite{Buras:2018hze}.
\end{abstract}

\setcounter{page}{0}
\thispagestyle{empty}
\newpage



\section{Introduction}

The ratio
$\epe$  measures the size of the direct CP violation in $K_L\to\pi\pi$ decays
relative to the indirect one described by $\varepsilon_K$ and 
is very sensitive to new sources of CP violation. It has
been measured already 16 years ago with
the experimental world
average from NA48 \cite{Batley:2002gn} and KTeV \cite{AlaviHarati:2002ye, 
Abouzaid:2010ny} collaborations given by
\begin{align}
  \label{eq:epe:EXP}
  (\epe)_\text{exp} & = (16.6 \pm 2.3) \times 10^{-4}\,.
\end{align}

In the Standard Model (SM)
$\varepsilon^\prime$ is governed by a positive contribution from QCD penguins  but receives also an important  contribution from electroweak (EW) penguins that  enter $\varepsilon^\prime$  with the opposite sign.  This is best seen in an analytic formula for $\epe$  
\cite{Buras:2015yba}
\begin{equation}
\left(\frac{\varepsilon'}{\varepsilon}\right)_{\text{SM}} =  10^{-4} \biggl[
\frac{\IM\lambda_{\rm t}}{1.4\cdot 10^{-4}}\biggr] \left[\,
a\big(1-\hat\Omega_{\rm eff}\big) \big(-4.1 + 24.7\,\bsi\big) + 1.2 -
10.4\,\bei \,\right] ,
\label{AN2015}
\end{equation}
where the contributions from the dominant QCD penguin $(Q_6)$ and the dominant EW penguin $(Q_8)$ are the ones proportional 
to the hadronic parameters $\bsi$ and $\bei$, respectively. Next
\be
\IM\lambda_{\rm t}=\IM ({V_{td}V_{ts}^*})= \vub\vcb \sin\gamma
\ee
and \cite{Cirigliano:2003nn,Cirigliano:2003gt,Bijnens:2004ai,Buras:2015yba}
\be\label{Omega}
a=1.017, \qquad \hat\Omega_{\rm eff} = (14.8 \pm8.0)\times 10^{-2}\,.
\ee
$\vub$ and $\vcb$ are the elements of the CKM matrix and $\gamma$ is the known 
angle in the unitarity triangle.
The parameters $a$  and $\hat\Omega_{\rm eff}$  represent isospin breaking corrections \cite{Cirigliano:2003nn,Cirigliano:2003gt,Bijnens:2004ai}.  See latter papers  and \cite{Buras:2015yba} for details.
$\hat\Omega_{\rm eff}$ differs from $\Omega_{\rm eff}$ in \cite{Cirigliano:2003nn,Cirigliano:2003gt} that includes EW penguin  contributions 
to ${\IM} A_0$. 
We find it more natural to calculate ${\IM} A_0$ including both QCD and EW penguin contributions as this allows to keep track of new physics (NP) contributions.

Setting all parameters, but $\bsi$ and $\bei$, at their central values we find
\begin{equation}
\left(\frac{\varepsilon'}{\varepsilon}\right)_{\text{SM}}  =  10^{-4} 
\left[\, -3.6 + 21.4\,\bsi + 1.2 -
10.4\,\bei \,\right].
\label{NUM2015}
\end{equation}
The first term ($-3.6$) comes from $(V-A)\times(V-A)$ QCD penguins and the third one ($1.2$) from  $(V-A)\times(V-A)$ EW penguins. These two contributions have
been determined in \cite{Buras:2015yba} from the experimental values of ${\RE} A_0$ and ${\RE} A_2$ with $A_{0,2}$ being isospin amplitudes. In doing this one
makes a plausible assumption   that the real parts of these amplitudes and 
the related $\Delta I=1/2$ rule can be 
properly described within the SM. The main uncertainty in $\epe$ resides 
then in the contribution of $(V-A)\times(V+A)$ QCD penguin operator $Q_6$ and the one  from  $(V-A)\times(V+A)$ EW penguin operator $Q_8$ that enters $\epe$ with a negative sign.

In the strict large $N$ limit,  
$N$ being the number of colours, one finds  \cite{Buras:1985yx,Bardeen:1986vp,Buras:1987wc}
\be\label{LN1}
\bsi=\bei=1\,, \qquad {\rm (large~N~Limit)}\,.
\ee
Inserting these values into (\ref{NUM2015}) we find
\be\label{epeLN}
\left(\frac{\varepsilon'}{\varepsilon}\right)_{\text{SM}} = 8.6\times 10^{-4},\qquad ({\rm using~~ (\ref{LN1}})),
\ee
roughly by a factor of two below the central experimental value in (\ref{eq:epe:EXP}). In fact this was a typical value obtained by the Rome \cite{Ciuchini:1995cd} and Munich \cite{Bosch:1999wr} groups in the 1990s, which  used the values in (\ref{LN1}). 
However, the authors in \cite{Antonelli:1995gw,Bertolini:1995tp,Pallante:1999qf,Pallante:2000hk,Buchler:2001np,Buchler:2001nm,Pallante:2001he}  using 
the ideas from Chiral Perturbation Theory ($\chi$PT) made a strong claim 
that final state interactions (FSI)  enhance $\bsi$ above unity and suppress 
$\bei$ below it so that the SM value for $\epe$ according to them 
is fully consistent with experiment.
Albeit only a very inaccurate value $(17\pm9)\times 10^{-4}$ \cite{Pallante:2001he}   could be obtained. As there are some different views in the literature what is meant by 
FSI,  let me call  the effects pointed out in these papers simply {\em pion loops}.
In fact this is the wording used in recent papers by some of these authors.
More about it later.

Therefore, I suspect, that it was a great surprise, in particular for $\chi$PT experts, when in 2015 the RBC-UKQCD collaboration  \cite{Bai:2015nea,Blum:2015ywa} presented their first results for $K\to\pi\pi$ hadronic matrix 
elements. From their results 
one could extract values of $\bsi$ and $\bei$ for $\mu=1.53\gev$. They 
 are  \cite{Buras:2015yba}
\be\label{B8LAT0}
\bsi=0.57\pm 0.19,\qquad B_8^{(3/2)}=0.76 \pm 0.05\,,\qquad  {\rm (RBC-UKQCD)}.
\ee

Inserting the central values into our formula (\ref{NUM2015}) we find
\be
\left(\frac{\varepsilon'}{\varepsilon}\right)_{\text{SM}} = 1.9\times 10^{-4},
\ee
that is one order of magnitude below the experimental value.
This is in fact the central value for $\epe$ in \cite{Buras:2015yba} which including various uncertainties found
\begin{align}
  \label{eq:epe:LBGJJ}
  (\epe)_\text{SM} & = (1.9 \pm 4.5) \times 10^{-4},\qquad {\rm (BGJJ)}\,.
\end{align}
 This result has  been  confirmed within the uncertainties in \cite{Kitahara:2016nld}
\begin{align}
  \label{KNT}
  (\epe)_\text{SM} & = (0.96 \pm 4.96) \times 10^{-4},\qquad {\rm (KNT)}\,.
\end{align}

It should be stressed that the RBC-UKQCD lattice collaboration, calculating directly hadronic matrix elements of all operators, but not including isospin breaking (I.B.) effects, found 
\cite{Blum:2015ywa, Bai:2015nea}
\begin{align}
  \label{eq:epe:LATTICE}
  (\epe)_\text{SM} & = (1.38 \pm 6.90) \times 10^{-4},\qquad {\rm (RBC-UKQCD)}.
\end{align}
The larger error than in (\ref{eq:epe:LBGJJ}) and (\ref{KNT}) is related 
to the fact that RBC-UKQCD collaboration did not use experimental data for ${\RE} A_0$ and ${\RE} A_2$ for the extraction of the $(V-A)\times(V-A)$ contributions. Otherwise the error in (\ref{eq:epe:LATTICE}) would be smaller. The comparison 
of RBC-UKQCD value, that does not include I.B. effects, with the values in 
(\ref{eq:epe:LBGJJ}) and (\ref{KNT}) would seemingly indicate that such effects 
are small. But such a conclusion would be wrong. I.B. effects definitely 
suppress significantly $\epe$, as we have seen in (\ref{AN2015}), but the first negative term 
in this formula, that is dominated by the QCD penguin operator $Q_4$, is found by RBC-UKQCD to be significantly larger in magnitude than extracted from the experimental value of ${\RE} A_0$. This difference suppresses $\epe$ at a similar level as I.B. effects. I expect that future RBC-UKQCD calculations will result in the value of the first term in (\ref{AN2015}) that is close to the one used by us.

At a flavour workshop in Mainz in January 2016 two important $\chi$PT experts 
Gilberto Colangelo and Toni Pich expressed serious doubts about the RBC-UKQCD result in (\ref{eq:epe:LATTICE}), because the $(\pi\pi)_I$ 
 phase shift $\delta_0\approx (24\pm 5)^\circ$ obtained by RBC-UKQCD disagreed 
 with  $\delta_0\approx 34^\circ$ obtained by combining dispersion 
theory with experimental input \cite{Colangelo:2001df}.

This criticism appeared 
in print one year ago \cite{Gisbert:2017vvj} and in two recent conference proceedings \cite{Gisbert:2018tuf,Gisbert:2018niu}. 
It is in line with the one expressed many years ago in \cite{Antonelli:1995gw,Bertolini:1995tp,Pallante:1999qf,Pallante:2000hk,Buchler:2001np,Buchler:2001nm,Pallante:2001he}, but one should realize that with $\delta_0\approx 24^\circ$
a big portion of FSI has been taken into account already in (\ref{eq:epe:LATTICE}). From my point of 
you it is unlikely that increasing $\delta_0$ up to its dispersive  value 
$\delta_0\approx 34^\circ$ will shift $\epe$ upwards by one order of magnitude. Recently a new 
result for $\delta_0$ has been presented by the RBC-UKQCD collaboration that 
with $\delta_0=(30.9\pm 3.4)^\circ$ is within $1\sigma$ from its dispersive
 value\footnote{See talks by Ch.Kelly and T. Wang at Lattice 2018.}. Whether this change has a crucial impact on $\epe$ as expected in  \cite{Gisbert:2017vvj,Gisbert:2018tuf,Gisbert:2018niu}  remains to be seen but should be known within the  next six months.

\section{The DQCD View}\label{sec:DQCD}
Motivated by the RBC-UKQCD results in (\ref{B8LAT0}), 
 Jean-Marc G{\'e}rard and myself calculated already in July 2015 $1/N$ corrections to the large $N$ limit in  (\ref{LN1}) \cite{Buras:2015xba}. These corrections, loop corrections in the meson theory with a {\em physical cut-off} $\Lambda\approx 0.7\gev$, are the leading non-factorizable corrections to hadronic matrix elements 
of $Q_6$ and $Q_8$. Two main results in this paper are:
\begin{itemize}
\item
Realization that the large-$N$ result is not valid at scales $\ord(1\gev$), as
assumed in all papers before \cite{Buras:2015xba}, but at much lower scales 
$\ord(m_\pi^2)$. In order to find it out one has to calculate these loops. In 
the large $N$ limit one cannot determine the scale in $\bsi$ and $\bei$ and 
as for $\mu\ge 1\gev$ the $\mu$ dependence of these parameters is weak \cite{Buras:1993dy}, without
knowing $1/N$ corrections it was useful to neglect this dependence.
\item
Calculation of $\bsi$ and $\bei$ at scales $\ord(1\gev)$ by performing the meson evolution from  the low factorization scale $\ord(m_\pi^2)$ to the physical 
cutoff $\Lambda$ of DQCD with the result
\be\label{bsibei}
\bsi\le 0.54, \qquad \bei=0.80\pm 0.10,
\ee
in perfect agreement with the RBC-UKQCD result in (\ref{B8LAT0}). Explicitly 
 we found
\be
\bsi=1-0.66\,\ln(1+\frac{\Lambda^2}{\tilde m_6^2}), \qquad
\bei =1-0.17\, \ln(1+\frac{\Lambda^2}{\tilde m_8^2}),
\ee
with pseudoscalar mass scale parameters $m_{6,8}$ bounded necessarily by the cut-off $\Lambda$: $\tilde m_{6,8}\le \Lambda$. The upper bound for $\bsi$ is 
obtained by setting $\Lambda=\tilde m_6$.
\end{itemize}

As already mentioned, for scales above $1\gev$ both parameters decrease very slowly. This is known from the 1993 analysis in \cite{Buras:1993dy} but as seen in Figs. 11 and 12 of that paper $\bsi$ decreases faster with increasing scale than $\bei$ in accordance with the pattern at low scales found by Jean-Marc and 
myself. This can also be shown analytically \cite{Buras:2015xba}.
Unfortunately, not knowing $1/N$ corrections to $\bsi$ and $\bei$ in 1993, both parameters have been set  at $\mu=m_c$ in \cite{Buras:1993dy} to unity, which is clearly wrong.

While it is possible that the bound on $\bsi$ could be violated by $1/N^2$ 
corrections and other effects not taken by us into account, one should notice that  with only pseudoscalars included in the loops, the cut-off $\Lambda$ has to  be chosen below $1\gev$ so that these 
 omitted effects, even if they would increase 
 $\bsi$, could still be compensated by the running to 
higher scales that are explored by lattice QCD.

The second result is in my view very important for the following reason. There
is no other lattice collaboration calculating $\bsi$ and $\bei$ at present, so that 
in the lattice world the result of the RBC-UKQCD collaboration for $\epe$ cannot be tested
at present. As we will see in Section~\ref{sec:GP}, $\chi$PT by itself has no means 
to verify or disprove the RBC-UKQCD results for $\bsi$ and $\bei$. Therefore, it is really important that DQCD provides the insight in the values of these 
parameters, which clearly cannot be provided by a purely numerical method like 
lattice QCD. According to our analysis in \cite{Buras:2015xba},
the main QCD dynamics behind the lattice values in (\ref{B8LAT0}) is the
meson evolution, analogous to the well known quark-gluon evolution at short 
distance scales.

Before I write a few lines about FSI in DQCD, let me just mention that the
importance of the meson evolution is also seen in the evaluation of BSM matrix elements relevant for $K^0-\bar K^0$ mixing. Indeed the
RBC-UKQCD collaboration working at $\mu=3\gev$ finds for four BSM parameters 
\cite{Garron:2016mva,Boyle:2017skn,Boyle:2017ssm}
\be\label{L23}
 B_2=0.488(7)(17), \qquad  B_3=0.743(14)(65),
\ee
and
\be\label{L45}
B_4=0.920(12)(16),\qquad B_5=0.707(8)(44),
\ee
with the first error being statistical and the second systematic. Similar 
results are obtained by ETM  \cite{Carrasco:2015pra} and 
SWME \cite{Jang:2015sla} collaborations, although the values for 
$B_4$ and $B_5$ from the ETM collaboration are visibly below the ones from RBC-UKQCD given above: $B_4=0.78(4)(3)$ and $B_5=0.49(4)(1)$.
Except for $B_4$ all values differ significantly from unity prohibiting the use 
of the vacuum insertion method.

In DQCD in the large $N$ limit, in which only {\em factorizable} contributions are
present,  one finds on the other hand \cite{Buras:2018lgu}
\be\label{B25}
B_2=1.20, \qquad  B_3=3.0\,, \qquad 
B_4=1.0, \qquad  B_5=0.2 \qquad ({\rm large~N~limit})\,.
\ee
These results differ drastically from the lattice results given above except 
for $B_4$. For instance $B_3$ is by a factor of four larger than the lattice result and $B_5$ by a factor of 3.5 smaller. But they apply 
to $\mu=\ord(m_\pi)$ while the lattice results where obtained at $\mu=3\gev$.
Very importantly, as demonstrated in \cite{Buras:2018lgu}, the meson evolution to scales $\ord(1\gev)$ followed by the usual quark-gluon evolution, allows again to understand the pattern of lattice values in (\ref{L23}) and (\ref{L45}) and agrees with them within $5-20\%$ depending on the considered parameter. Again the meson evolution and not the quark-gluon 
evolution is responsible for this insight. Moreover, this insight in the 
lattice results has been obtained basically without any free parameters, except 
for the cut-off $\Lambda$ which being physical is in any case approximately known and in the ballpark of $0.7\gev$, if only pseudoscalars are present in the loops as done in \cite{Buras:2018lgu}. Moreover, our calculation has been done in 
the chiral limit and we expect that going beyond this limit the agreement 
with RBC-UKQCD values would be better. I should stress that our goal was by no means to compete or even verify that RBC-UKQCD results are correct because this 
has been done already by two other collaborations. The main goal of our paper, 
unfortunately missed by {\em all} referees of our paper in JHEP and EPJC, was to gain 
the following  important lesson from this analysis:
\begin{itemize}
\item
Working in the strict large $N$ limit in non-leptonic Kaon decays misses a 
very important QCD dynamics represented by meson evolution so that it is mandatory to take it into account in any sensible phenomenology. As demonstrated in 
\cite{Buras:2018lgu} this dynamics is hidden in lattice QCD results for $K^0-\bar K^0$ mixing but as we will see in Section~\ref{sec:GP} it is missed in the $\chi$PT framework.
\end{itemize}

Now the controversial  FSI are absent in the $K^0-\bar K^0$ mixing but 
are certainly present in $K\to\pi\pi$ decays and as claimed by $\chi$PT researchers 
they are responsible for a significant enhancement of $\epe$ over the RBC-UKQCD results 
as described above. Jean-Marc and myself looked at this issue  
in \cite{Buras:2016fys} and reached the following conclusions:
\begin{itemize}
\item
In $\chi$PT, without additional dynamics, it is impossible to separate the current-current operator $Q_2-Q_1$,  
responsible dominantly for the $\Delta I=1/2$ rule, from the QCD penguin operator $Q_6$ that is irrelevant for this rule at $\mu=m_c$ and amounts to at most 
$15\%$ correction at $\mu=1\gev$. Therefore claiming that 
 pion dynamics, which could indeed provide an enhancement of $\RE A_0$, enhances 
the matrix element of $Q_6$ at the same level as  $\RE A_0$ is questionable.
\item
Our claim is that the suppression of $\epe$ by the meson evolution, being of $\ord (p^0/N)$ in the case of the $Q_6$ operator, must be more important than the enhancement of $\epe$ by FSI that are $\ord(p^2/N)$.
\end{itemize}

Finally, let me react to the statements made in \cite{Gisbert:2017vvj,Gisbert:2018tuf} that in all papers I was involved in, all absorptive parts 
have been set to zero and that the analysis in \cite{Buras:2015yba} leading to (\ref{eq:epe:LBGJJ}) was 
done in the context of DQCD. This is fake news. The analysis in  \cite{Buras:2015yba} used entirely RBC-UKQCD results that include absorptive parts as 
mentioned above.

What has been obtained in DQCD is the  upper bound \cite{Buras:2015xba,Buras:2015yba}
\be
  \label{DQCDA}
  (\epe)_\text{SM}  \le (6.0 \pm 2.4) \times 10^{-4}\,,\qquad ({\rm DQCD})
\ee
that follows from $\bsi\le \bei$ with $\bei$ taken from RBC-UKQCD collaboration. It includes other uncertainties, but not the NNLO QCD corrections which provide a downward shift of $\epe$ by $1-2$ units as we will discuss now.

\section{Few messages from short distance}
All present analyses of $\epe$ are based on the Wilson coefficients 
of QCD and EW penguin operators evaluated
already 25~years ago at NLO in \cite{Buras:1991jm, Buras:1992tc, Buras:1992zv,
Ciuchini:1992tj, Buras:1993dy, Ciuchini:1993vr}. While at this level 
some removal of unphysical renormalization scheme and scale dependences present
at LO takes place in the case of QCD penguins, as pointed out in \cite{Buras:1999st}, this is not the case
for EW penguin contributions. In fact, as
 emphasized in \cite{Aebischer:2018csl},
all the existing estimates of $\epe$ at
NLO suffer from short-distance renormalization scheme uncertainties
in EW penguin contributions and also scale uncertainties in 
$m_t(\mu)$ that were practically removed in the NNLO matching at the electroweak scale already 20 years ago \cite{Buras:1999st}. In the naive dimensional regularization
(NDR) scheme, used in all recent analyses, these corrections enhance the
electroweak penguin contribution by roughly $16\%$, thereby leading to a {\em negative}
shift of $-1.3\times 10^{-4}$ in $\epe$, decreasing further its  value, similar to isospin breaking effects.
This could appear small in
view of other uncertainties. However, on the one hand, potential scale and
renormalization scheme uncertainties have been removed in this manner and on
the other hand, one day such corrections could turn out to be relevant. Finally,
the fact that this correction further decreases $\epe$ within the SM gives
another motivation for the search for NP responsible
for it. 

The world is now waiting for the final results of NNLO QCD corrections to 
QCD penguin contributions to $\epe$. Preliminary results appeared already in
\cite{Cerda-Sevilla:2016yzo,Cerda-Sevilla:2016tpw,Cerda-Sevilla:2018hjk} and 
looking at Fig.\,1 in \cite{Cerda-Sevilla:2018hjk} we observe that the 
scale uncertainties in $\epe$ have been practically eliminated and that 
these corrections slightly suppress $\epe$ further. Yet, in order to be able 
to put this 
observation on solid grounds, we badly need final result from these impressive
efforts.

We now turn to the highlights of this presentation.

\section{The anatomy of the Gisbert-Pich plot}\label{sec:GP}
In a detailed review \cite{Gisbert:2017vvj} and two conference proceedings \cite{Gisbert:2018tuf,Gisbert:2018niu} Gisbert and Pich presented 
their view on the present status of $\epe$ within the SM. They work within the 
context of $\chi$PT which at its basis uses low energy symmetries of QCD. However, 
it should be realized that  without  any additional dynamical input from lattice QCD or large $N$ approach, like DQCD,  hadronic matrix elements relevant for $\epe$ cannot be calculated in $\chi$PT. 

At this point it should be stressed that DQCD is not a {\em constrained} $\chi$PT,
as stated by one referee of our papers, and 
anybody who thinks like this has missed completely the main ideas of DQCD\footnote{For a pedagogical introduction see \cite{Gerard:1990dx,Fatelo:1994qh} and a more recent 
reviews are \cite{Buras:2014maa,Buras:2018hze}.}.  
The differences between DQCD and $\chi$PT have been spelled out in many of our papers 
and I will not repeat them here. Probably the best comparison is given in Section 3.3 of \cite{Buras:2014maa}. Let me just mention two points:
\begin{itemize}
\item
Even if in both DQCD and $\chi$PT meson loops are involved,
DQCD uses a {\em physical} cut-off in the calculation of these loops and thereby allows to restrict the loop momenta to the region in which the truncated meson theory is valid. In this manner one achieves a clear separation between long and short distance contributions which in turn allows 
 through the meson evolution a
proper matching of hadronic matrix elements with short distance dynamics. 
On the other hand  $\chi$PT uses dimensional regularization which results in counter terms represented by low energy constants which can either be extracted from experiment or calculated by lattice QCD. As we will see soon, a proper matching with short distance physics is very difficult in this approach.
\item
While in DQCD the $Q_6$ operator can easily be separated from current-current operators, as discussed in \cite{Buras:2016fys}, this not possible in $\chi$PT unless some dynamics like QCD at large $N$ is used. See our discussion in Section~\ref{sec:DQCD}.
\end{itemize}

\begin{figure}[!tb]
 \centering
\includegraphics[width = 0.90\textwidth]{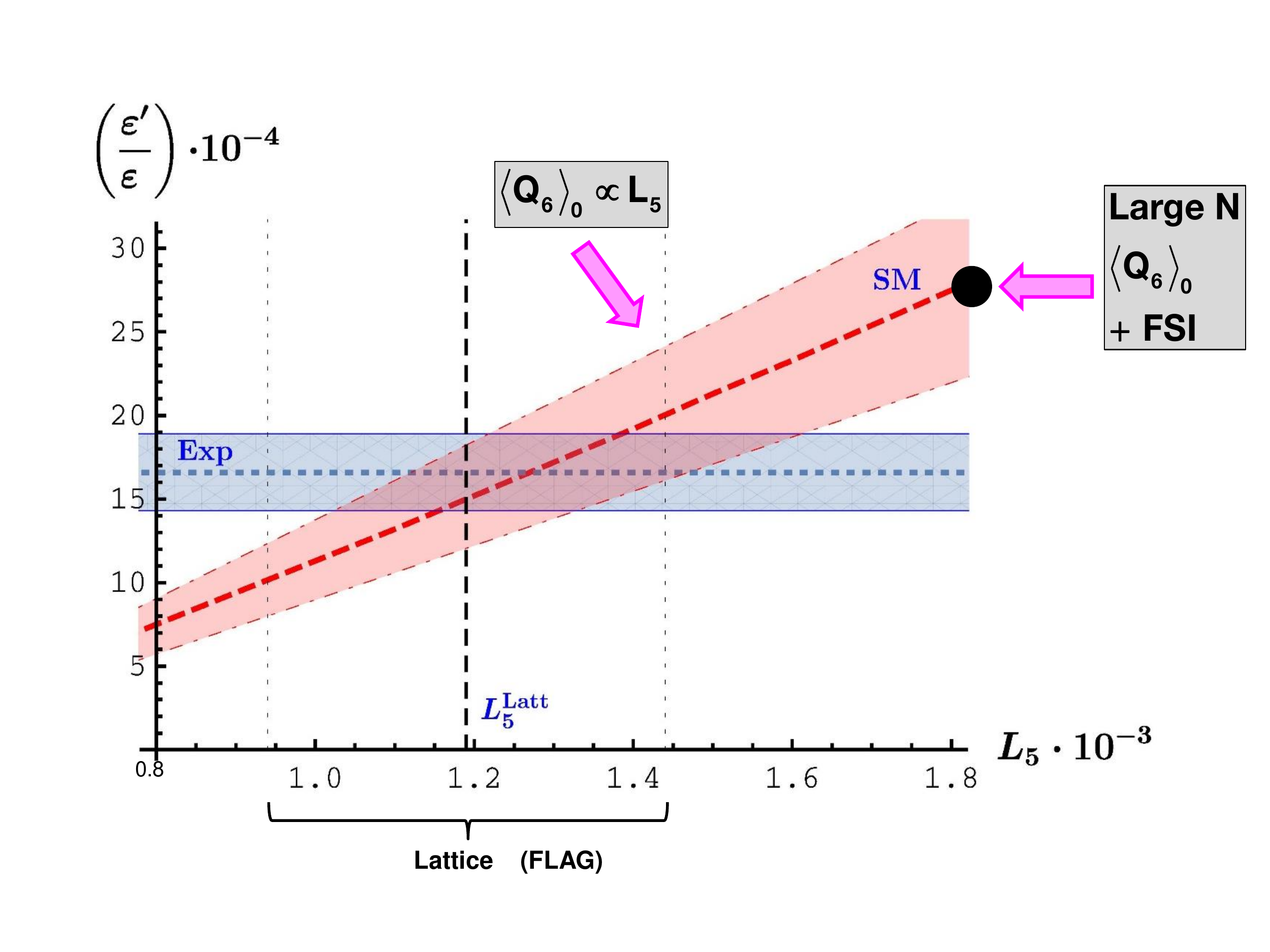}
\caption{ The Gisbert-Pich plot.
}\label{fig:GP}~\\[-2mm]\hrule
\end{figure}

 Central for \cite{Gisbert:2017vvj} and the subsequent two papers is the plot in 
Fig.~\ref{fig:GP}, to be termed GP-plot in what follows. It shows the 
dependence of $\epe$ on the low-energy constant $L_5$. It is obtained by finding the $\chi$PT realization of the $\Delta S=1$ effective Lagrangian in the large $N$ 
limit which allows to find the coupling $g_8$ in this limit as given in 
(47) of  \cite{Gisbert:2017vvj}. It is a linear combination of contributions 
of current-current operators and the penguin operators $Q_4$ and $Q_6$. The one 
involving $Q_6$ is proportional to $L_5$ which is varied in Fig.~\ref{fig:GP}. 
In addition the authors include pion loops representing FSI but the effect 
of meson evolution is absent in their calculation as the inclusion of it
is beyond the $\chi$PT framework.

For our purposes 
it will be useful to find an analytic formula for the central dotted red line in Fig.~\ref{fig:GP}.  It reads
\be\label{AJBL5}
(\epe)_{\rm SM}= (19.2 \, L_5-7.8)\cdot 10^{-4}.
\ee

In the absence of the meson evolution, which has nothing to do with pion loops 
of these authors, one has $L_5=1.84\cdot 10^{-3}$ which corresponds to the 
large $N$ limit of $\bsi$ in (\ref{LN1}), that is $\bsi=1$. This 
implies SM value of 
$\epe$ as large as $27.5\cdot 10^{-4}$, roughly by a factor of three larger 
than in (\ref{epeLN}). Indeed the impact of pion loops on $\epe$, as calculated in  \cite{Gisbert:2017vvj}, is very large.  But, in spite of the fact that (47) in  \cite{Gisbert:2017vvj} is valid only in the large $N$ limit, eventually the authors decide to use the current FLAG compilation for $L_5$ which reads
\be\label{L5LATT}
L^{\rm Latt}_5=(1.19\pm 0.25)\cdot 10^{-3},
\ee
and through (\ref{AJBL5}) implies the $1\sigma$ range
\be
10.2\times 10^{-4}\le (\epe)_{\rm SM} \le 19.8 \times 10^{-4}\,.
\ee
As $L_5$ from FLAG surely includes some $1/N$ corrections, and (47) in  \cite{Gisbert:2017vvj}  has 
been obtained in the large $N$ limit, it is not evident 
that such a procedure is really self-consistent. But let us not enter this 
issue in our Christmas story.

In any case, after the inclusion of other uncertainties, their final result
for $(\epe)_{\rm SM}$ reads
\be\label{GPSM}
(\epe)_{\rm SM}=(15\pm7)\cdot 10^{-4}\,.
\ee

This according to the authors of \cite{Gisbert:2017vvj} is the present status of $\epe$ within the 
SM. I certainly disagree with it. It can only be their 
view what $\epe$ in the SM is. As we will see soon, one cannot even state that
this is the view of all $\chi$PT experts. But let us first see what are the 
implications of the result in (\ref{GPSM}) taken at face value:
\begin{itemize}
\item
There is no $\epe$ anomaly.
\item
There is still a large room for NP contributions, as even within $1\sigma$, values
as low as $8\cdot 10^{-4}$ and as high as $22\cdot 10^{-4}$ are allowed. 
\item 
Consequently this result is  not motivating for NP searches. NP 
enhancing significantly $\epe$ and also NP suppressing it even by a factor of
two are both still allowed. 
\end{itemize}

\begin{figure}[!bt]
 \centering
\includegraphics[width = 0.90\textwidth]{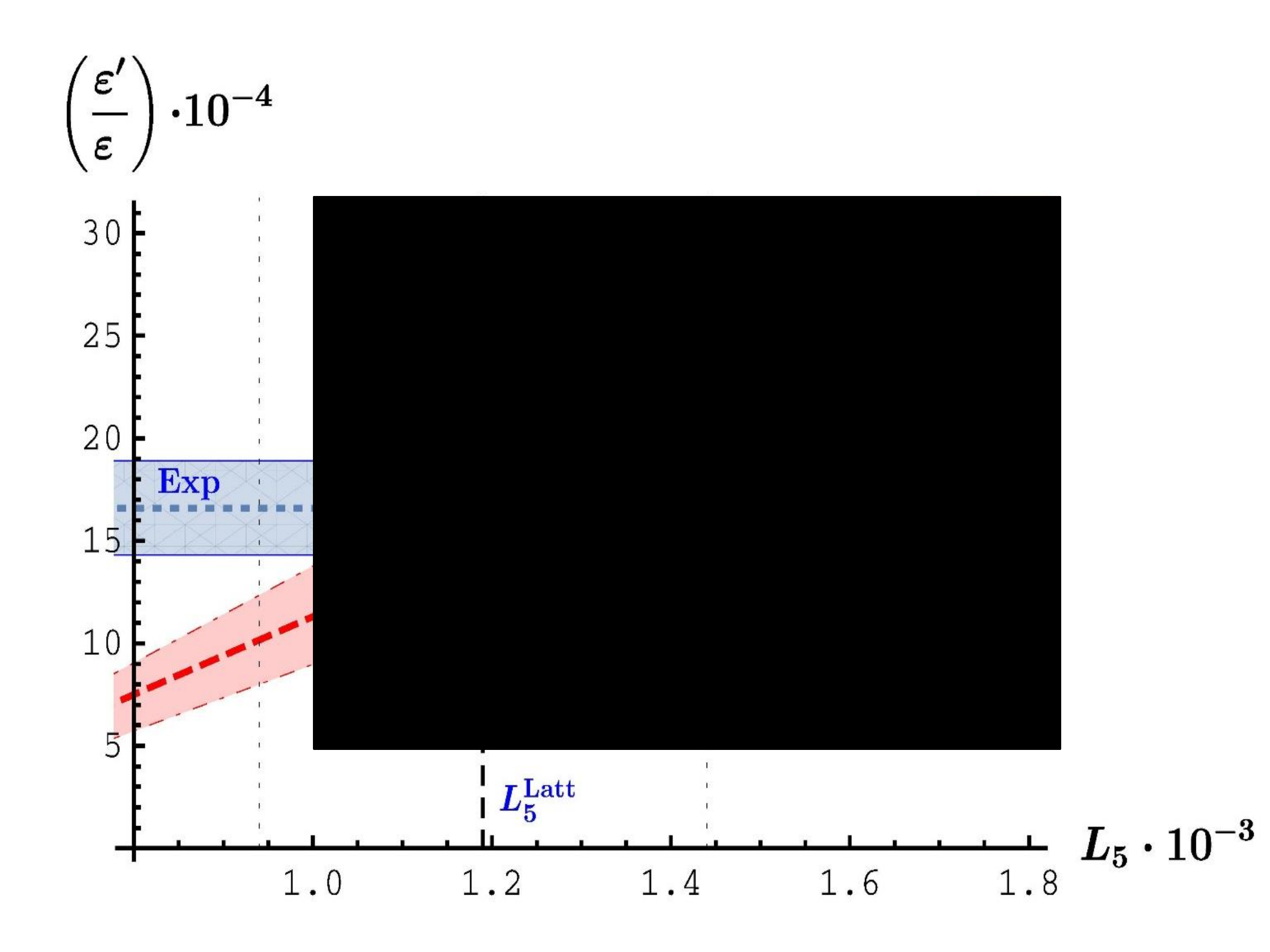}
\caption{The impact of meson evolution on the Gisbert-Pich plot.
}\label{fig:AJB}~\\[-2mm]\hrule
\end{figure}

Fortunately, the authors are aware of this problematic and our
statements are supported by a sentence of Gisbert and Pich in 
\cite{Gisbert:2017vvj} which they repeated in \cite{Gisbert:2018tuf}:\\

{\em Our dominant uncertainty reflects our ignorance about $1/N$-suppressed  contributions that we have missed in the matching process.}\\

In my language this refers to the absence of meson evolution in their
calculation which they want to replace  somehow by using the FLAG value 
of $L_5$ that is rather uncertain. 

Let me then perform the following exercise. I take at {\em face value}  the GP calculation
 of pion loops that with the large $N$ value for $L_5$ resulted in a large 
value of $\epe$. But as we have seen in Section~\ref{sec:DQCD}, in the case of $K^0-\bar K^0$ mixing  our calculation of hadronic matrix elements in the large-$N$ limit
totally misrepresented their values obtained by three  lattice QCD groups. 
I definitely do not need the FLAG value for $L_5$ as DQCD is not ignorant 
about $1/N$ suppressed contributions that are included by means of the 
meson evolution in the matching process.
 I can therefore find the 
impact of the meson evolution on the plot in Fig.~\ref{fig:GP} without any 
help from lattice QCD. It is shown 
in Fig.~\ref{fig:AJB}. Not much is left from the GP-plot as a large part of this plot is eliminated by the 
meson evolution. Few comments are in order.
\begin{itemize}
\item
Our result $L_5\le 1.0 \cdot 10^{-3}$, that follows from the upper bound on $\bsi$ in 
(\ref{bsibei}) and the large $N$ value $L_5=1.84\cdot 10^{-3}$ corresponding to $\bsi=1$, is consistent with the FLAG value of $L_5$, but is significantly below FLAG's central value.
\item
In view of my comments on the size of the impact of FSI on $\epe$, as calculated 
presently in $\chi$PT,  in Section~\ref{sec:DQCD}, I expect that in reality 
the black region is shifted significantly to the left. 
\item
But even with the result in  Fig.~\ref{fig:AJB}, adding NNLO QCD corrections 
to EW-penguins, absent in the GP-plot, I end up with an upper bound
\be\label{AJBbound}
(\epe)_{\rm SM}< 11\cdot 10^{-4},
\ee
which in my view is conservative.
\end{itemize}

Thus even including GP's pion loops, the meson evolution implies a value 
of $(\epe)_{\rm SM}$ significantly below the data. This eliminates NP scenarios
which suppress $\epe$. I leave it to the readers to decide whether this result can be considered as a hint for an $\epe$ anomaly or not.

But now comes a surprise. As discovered by Jean-Marc, in a 2014 paper two 
important $\chi$PT experts, Bijnens and 
Ecker, making apparently a more sophisticated analysis of $L_5$ than done by 
FLAG, found  the range $0.5\le L_5\le 1.0$ \cite{Bijnens:2014lea}. This 
through (\ref{AJBL5}) implies the range
\be
1.8\times 10^{-4}\le (\epe)_{\rm SM} \le 11.4 \times 10^{-4}\,,\qquad ({\rm BE})
\ee
in perfect agreement with the bound in (\ref{AJBbound}) and the very low 
value for $(\epe)_{\rm SM}$ predicted within DQCD. It is beyond my skills 
to judge whether the FLAG number or the number following from \cite{Bijnens:2014lea}, is correct, but
it appears that the uncertainty in the estimate of $(\epe)_{\rm SM}$ by $\chi$PT 
is even larger than shown to us by Gisbert and Pich.

\section{Congratulations to NA48 and KTev}
\begin{figure}[!bt]
 \centering
\includegraphics[width = 0.90\textwidth]{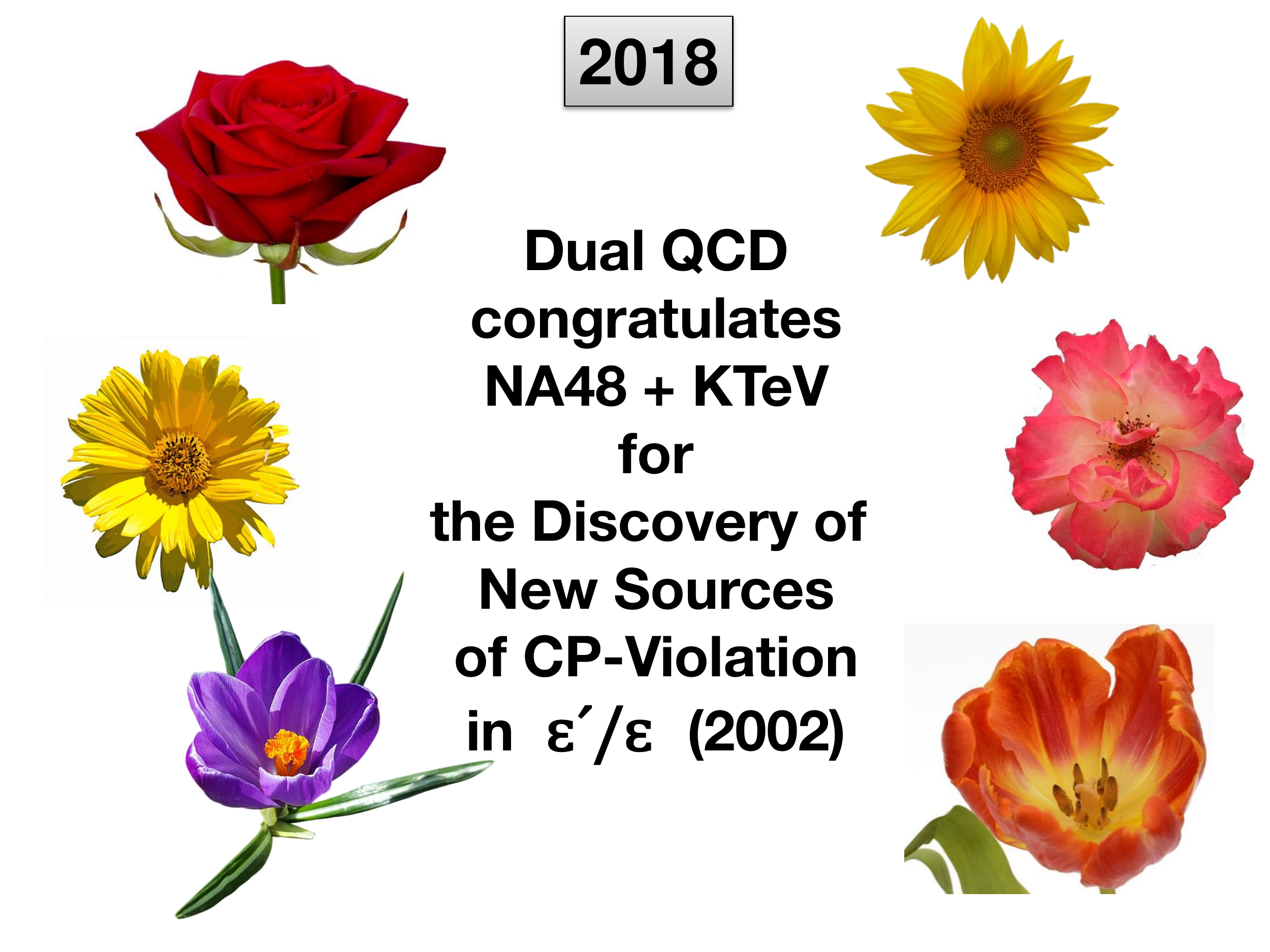}
\caption{Congratulations to NA48 and KTeV.
}\label{fig:flowers}~\\[-2mm]\hrule
\end{figure}

I have reached my goal now. On the basis of my work with Jean-Marc G{\'e}rard and the findings  above I
am ready to sign the statement in Fig.\,\ref{fig:flowers}.

However, similar to the case of $R_K$ and $R_{K^*}$ anomalies hinted by the LHCb data, which require the confirmation by Belle II, the $\epe$-anomaly hinted by 
DQCD requires it's confirmation by lattice QCD before it is firmly established.
Moreover, while DQCD provides presently a rough range for NP contribution 
to $\epe$
\be
7\cdot 10^{-4}\le (\epe)_{\rm NP} \le 15\cdot 10^{-4},
\ee
one should hope that LQCD will provide in due time a much more narrow range 
that is crucial for NP searches.

I already discussed in detail why $\chi$PT is not in a position to congratulate 
NA48 and KTeV for this important discovery, so let me explain in a  few lines why 
this is also the case for lattice QCD. In fact none of the members 
of the RBC-UKQCD collaboration, I talked to, claims that their result implies NP physics and 
some are surprised that I am looking for it. Let me stress again, my motivations
 for looking for NP in $\epe$ are definitely not based on the results from RBC-UKQCD, but on our results in DQCD. Yet, it is gratifying that RBC-UKQCD gives 
some support for them. Now come few lines about lattice QCD in general.

Lattice QCD is like an experiment, you perform very sophisticated numerical calculations based on first principles of QCD, which eventually will tell us one day with high precision what is the value of $\epe$ in the SM. I admire the RBC-UKQCD  collaboration
for their efforts and also other collaborations, in particular those which 
calculate hadronic matrix elements and weak decay constants relevant for 
$B$ physics. They all should get required financial support to reach their
goals. Yet the following should be realized. 

The numerical values of various hadronic parameters lattice groups 
 present to us come 
from a {\em Blackbox} without any insight why these  values are what they 
are. For flavour phenomenology this insight is generally irrelevant. One can 
just use these numbers at face value hoping that they are right. Similar to experiments, one needs results from at least two lattice collaborations which agree with each other to be sure that they are correct. 
Moreover, 
as far as $B$-physics and charm physics is concerned, I do not know any method which would 
allow us to look inside the lattice Blackboxes. But in $K\to\pi\pi$ decays the 
situation is different.
We have DQCD, which indeed allows us to see, at least roughly, what happens inside these
Blackboxes. I just want to quote two examples:
\begin{itemize}
\item
The parameter $\hat B_K$ entering the phenomenology of $\varepsilon_K$ is $0.75$ in the large $N$ limit \cite{Buras:1985yx}.
It takes at most one minute for beginners and few seconds for flavour experts 
to find this out.
In DQCD, including $1/N$ corrections, it takes, say, a week to find 
 $\hat B_K=0.67\pm 0.07$ \cite{Bardeen:1987vg}, when only the
 octet of pseudoscalar mesons is included. It takes another few weeks 
to include lowest lying vector mesons which allows to increase the cut-off $\Lambda$
to $0.85\gev$ and to improve the matching with short distances to find 
$\hat B_K=0.73\pm 0.02$ \cite{Buras:2014maa}. Several lattice groups calculated $\hat B_K$ over 
 almost 20 years resulting in the latest FLAG value $\hat B_K=0.763\pm 0.010$.
I have asked many lattice experts what is their explanation of $\hat B_K$ 
being so close to its value in the large $N$ limit. I did not get any answer. In DQCD the 
answer is simple. The loops with pseudoscalars suppress $\hat B_K$, while 
the ones with vector mesons enhance it. But vector meson contributions are mass  suppressed relative to pseudoscalar ones, so that they loose in this competition resulting in $\hat B_K < 0.75$. This argument is consistent with a more 
detailed analysis of Jean-Marc \cite{Gerard:2010jt}. Thus DQCD has still another prediction: FLAG's central value will go below 0.75 one day.
\item
The case of BSM parameters $B_{2-5}$ discussed already in Section~\ref{sec:DQCD}, in which DQCD explained the pattern of their values found by lattice QCD 
\cite{Buras:2018lgu}.  I challenge 
both lattice experts and $\chi$PT experts to explain this pattern.
\end{itemize}

Other insights provided by DQCD, like the one related to the $\Delta I=1/2$ rule, are described in  particular in \cite{Buras:2018hze,Buras:2014maa}. Yet, one of 
the referees of our analysis in \cite{Buras:2018lgu} stated that the insight 
 into the lattice values $B_{2-5}$, that we provided  in that paper, is of no interest to the flavour community. It could be. In the 
21st century possibly most people are only interested in numbers coming out from 
Blackboxes. But one should realize that this insight, involving four numbers, gives a very strong 
support to the relevance of the meson evolution, which represents the crucial QCD 
dynamics at scales below $1\gev$ and thereby gives an additional support to the $\epe$ anomaly. 

In this context I would like to mention our 2018 calculation of all BSM 
hadronic matrix elements contributing to $\epe$ \cite{Aebischer:2018rrz}. Also in this case 
the results in the large $N$ limit totally misrepresent the values of 
hadronic matrix elements at scales $\ord(1\gev)$ that we calculated in 
\cite{Aebischer:2018rrz} by performing the
 meson evolution. Similarly to the case of $B_K$, $\bsi$, $\bei$ and $B_{2-5}$,
the pattern of meson evolution agrees well with the quark-gluon evolution 
above $1\gev$. No lattice QCD results for these matrix elements are known 
and it will be interesting to see what lattice QCD will find for them one 
day.

All this shows that with the help of DQCD we can probe,  in the case of $K\to\pi\pi$, the dominant QCD effects 
in the Blackboxes of our lattice colleagues. As we have seen $\chi$PT has no means to do it. 
\section{Summary}
My Christmas story approaches the end. Let me make some final comments
\begin{itemize}
\item
Meson evolution and separation of $Q_6$ from $Q_2-Q_1$ is crucial for the correct estimate of $\epe$ within the SM. It is an important advantage of DQCD and 
lattice QCD over $\chi$PT. Lattice QCD experts are  not aware of the fact that 
they include meson evolution but we demonstrated this in several papers.
\item
Even with pion loops of  Gisbert and Pich, the inclusion of the meson evolution 
and NNLO QCD corrections to EW penguins results in  $(\epe)_{\rm SM}$ 
significantly below the data.  This is an important message for model
builders. NP has to provide an upward shift in $\epe$ in order to reproduce
data.
\item
The fate of $(\epe)_{\rm NP}$ depends now on the improved values for $\bsi$ 
and $\bei$ from lattice QCD, improved treatment of FSI and the inclusion 
of isospin breaking effects.
\item
The calculation of isospin breaking effects is presently led by $\chi$PT and 
in spite of my critical comments on some other aspects of this approach, 
I am sure that $\chi$PT will provide much more precise value than given in 
(\ref{Omega}) and this will happen likely  ahead of lattice QCD. In fact 
one of the first calculations of these corrections was done in 
\cite{Buras:1987wc} by Jean-Marc and myself with the 
result in the ballpark of $0.25$. This  is consistent with (\ref{Omega}), although a bit larger. 
The improvements on $L_5$ could also help $\chi$PT in predicting $(\epe)_{\rm SM}$.
\item
Last but not least the final result on NNLO QCD corrections to QCD penguin 
contributions will definitely reduce perturbative uncertainties. Also subleading
 contributions to NNLO QCD effects in EW penguins are still missing.
\end{itemize}

Personally, I am convinced about the presence of new physics in $\epe$  and prophesy that one day everybody will agree that \cite{Buras:2018wmb}
\be\label{AJBFINAL}
 (\epe)_{\rm SM}= (5\pm2)\cdot 10^{-4},\qquad (2026),
\ee
which  is also consistent with (\ref{DQCDA}). 
But this will take still several years. This number is based simply on the 
fact that in the large $N$ limit we have first $8.6\cdot 10^{-4}$. The suppression of $\epe$ through meson evolution will be partly compensated by FSI so 
that including these effects we will see numbers in the ballpark of 
$(7\pm 1)\cdot 10^{-4}$. Including NNLO QCD corrections will result in (\ref{AJBFINAL}). All this is very rough, but after spending over 30 years with $\epe$ 
these are my gut feelings what is its value within the SM. Most importantly 
they are supported by DQCD.

With this conviction in mind I plan to continue to search for NP responsible 
for $\epe$ anomaly as I did in the last 3.5 years. In this context let me just
mention our additional 2018  papers  that were not mentioned above \cite{Buras:2018evv,Aebischer:2018quc,Aebischer:2018csl}, 
       which are reviewed in 
\cite{Buras:2018wmb,Buras:2018hze,Aebischer:2018sst}. 
In particular in \cite{Buras:2018wmb}, which I plan to update, there is a table
 with references to  papers that performed several analyses of  $\epe$ in the context of various extensions of the SM.

 The next decade should be very exciting for flavour phenomenology, not only
 at Belle II \cite{Kou:2018nap} and LHCb \cite{Cerri:2018ypt},  but also 
generally at CERN where also ATLAS and CMS will contribute in an important 
manner  \cite{Cerri:2018ypt,CidVidal:2018eel}. I also expect that the results 
 on $\kpn$ from  NA62 collaboration at CERN
\cite{Ceccucci:2018bnw} and on $\klpn$ from  KOTO \cite{Ahn:2018mvc} at J-PARC,
when correlated with $\epe$, will allow a deep insight into possible NP at short
 distances \cite{Buras:2015jaq}. This insight will be enriched by 
 $(g-2)_\mu$ experiments at 
Fermilab and J-PARC and in the context of  various experiments probing charged lepton flavour violation 
at CERN, PSI, KEK and J-PARC. I also hope 
that my prophetic statements about $\epe$, very appropriate for a Christmas 
story, will be confirmed by several lattice QCD groups before my 80th birthday 
in 2026.

The discovery of new sources of CP violation by NA48 and KTeV collaborations is very important because we need them in order to explain our existence.   I really have no idea whether NP in $\epe$ is  responsible for our existence, not only because this is presently beyond my  skills but also because we did not yet identify what this NP is.
It could also happen that my strong believe in the presence of NP in $\epe$ 
will be refuted by lattice QCD one day. Yet, without the measurements of $\epe$ by NA48 and 
KTeV collaborations, all these discussions between RBC-UKQCD, $\chi$PT and DQCD 
 experts 
would be much less exciting and we should thank these two experimental groups 
for the result in (\ref{eq:epe:EXP}). I wish everybody, who still reads these 
lines, a Merry Christmas and a Happy 2019!

\section*{Acknowledgements}
 It is 
a pleasure to thank all my collaborators involved in the evaluation 
of $\epe$ in the SM and in various BSM scenarios. Particular thanks go to Jean-Marc for a wonderful 
collaboration over many years and for joined efforts to convince the 
community that Dual QCD is a successful low energy approximation of QCD. 
 Thanks are also due to Jason Aebischer for helping us in these efforts 
\cite{Aebischer:2018sst}. The improvements on V1 of this writing, suggested 
by Jean-Marc, Jason and Christoph Bobeth, are highly appreciated.
This research was  
supported by the DFG cluster of excellence ``Origin and Structure of the Universe''.

\renewcommand{\refname}{R\lowercase{eferences}}

\addcontentsline{toc}{section}{References}

\bibliographystyle{JHEP}
\bibliography{allrefs}
\end{document}